# On some difficulties in the addition of trapezoidal ordered fuzzy numbers

Anna Łyczkowska-Hanćkowiak[1], Krzysztof Piasecki[2]

**Abstract:** At the first, we revise the Kosinski's definition of the sum of ordered fuzzy numbers. The associativity of revised sum is investigated here. In addition, we show that the multiple revised sum of finite sequence of trapezoidal ordered fuzzy numbers depends on its summands ordering.

## 1. Introduction

Many mathematical applications require that finite multiple sum be independent on summands ordering. Any associative and commutative sum satisfies this property.

The sum of reals numbers is associative and commutative. The sum of fuzzy numbers (Dubois, Prade, 1979) is defined by means of the Zadeh Extension Principle (Zadeh, 1975 a, b, c). Also this sum is associative and commutative.

Kosiński and his co-writers (Kosiński et al, 2002; 2003; Kosiński, 2006) have introduced the concept of ordered fuzzy number as some extension of fuzzy number notion. The main goal of this paper is to investigate whether the finite sum of ordered fuzzy numbers is independent on summands ordering.

## 2. The basic notions

An imprecise number is a family of values in which each considered value belongs to it in a varying degree. A commonly accepted model of imprecise number is the fuzzy number (FN), defined as a fuzzy subset of the real line $\mathbb{R}$. The most general definition of FN is given as follows:

**Definition 1:** (Dubois, Prade, 1979) A fuzzy number is a fuzzy subset $\mathcal{S} \in \mathcal{F}(\mathbb{R})$, represented by its membership function $\mu_S \in [0,1]^{\mathbb{R}}$ satisfying the conditions:

$$\exists_{x\in\mathbb{R}}\, \mu_S(x) = 1; \tag{1}$$

$$\forall_{(x,y,z)\in\mathbb{R}^3}\colon\ x \le y \le z \Longrightarrow \mu_S(y) \ge \min\{\mu_S(x), \mu_S(z)\}. \ \square \tag{2}$$

In, addition, many applications of this definition requires some kind of membership function continuity. The set of all FN we denote by the symbol $\mathbb{F}$. Dubois and Prade (1978) first introduced the arithmetic operations on FN. These arithmetic operations are coherent with the Zadeh Extension Principle (Zadeh, 1975 a, b, c). The most popular type of FN is trapezoidal FN (TrFN).

**Definition 2:** For any nondecreasing sequence $\{a, b, c, d\} \subset \mathbb{R}$, the trapezoidal fuzzy number $\overleftrightarrow{Tr}(a, b, c, d)$ is defined by its membership functions $\mu_{Tr}(\cdot\, |a, b, c, d) \in [0,1]^{\mathbb{R}}$ as follows

[1] WSB University in Poznań, ul. Powstańców Wielkopolskich 5, 61-895 Poznań, Poland, E-mail:anna.lyczkowska.hanckowiak@wsb.poznan.pl
[2] Poznań University of Economics, Department of Investment and Real Estate, al. Niepodleglosci 10, 61-875 Poznań, Poland, E-mail: krzysztof.piasecki@ue.poznan.pl

$$\mu_{Tr}(x|a,b,c,d) = \begin{cases} 0 & x \notin [a,d] = [d,a] \\ \frac{x-a}{b-a} & x \in [a,b[ = ]b,a] \\ 1 & x \in [b,c] = [c,b] \\ \frac{x-d}{c-d} & x \in ]c,d] = [d,c[ \end{cases} \quad \square \qquad (3)^3$$

The concept of ordered fuzzy numbers (OFN) was introduced by Kosiński and his co-writers in the series of papers (Kosiński et al, 2002; 2003; Kosiński, 2006) as an extension of the concept of fuzzy numbers. The set of all OFN we denote by the symbol $\mathbb{K}$. The intuitive Kosiński's approach to the notion of OFN is very useful. For these reasons, below we present definition of trapezoidal OFN (TrOFN) which fully corresponds to the intuitive OFN definition by Kosiński.

**Definition 3:** For any monotonic sequence $\{a,b,c,d\} \subset \mathbb{R}$ trapezoidal ordered fuzzy number (TrOFN) $\overleftrightarrow{Tr}(a,b,c,d)$ is defined by its membership function given by (3). □

The condition $a < d$ fulfilment determines the positive orientation of TrOFN $\overleftrightarrow{Tr}(a,b,c,d)$. Positive oriented TrOFN is interpreted as such imprecise number, which may be bigger. For the case, the graph of the TrOFN $\overleftrightarrow{Tr}(a,b,c,d)$ membership function has an extra arrow denoting the positive orientation, which provides supplementary information. An example of such graph is presented in the Fig. 1a.

The condition $a > d$ fulfilment determines the negative orientation of OFN $\overleftrightarrow{Tr}(a,b,c,d)$. negative oriented trapezoidal OFN is interpreted as such imprecise number, which may be less. For the case, the graph of the TrOFN $\overleftrightarrow{Tr}(a,b,c,d)$ membership function has an extra arrow denoting the negative orientation. An example of such graph is presented in the Fig. 1b.

**Figure 1**

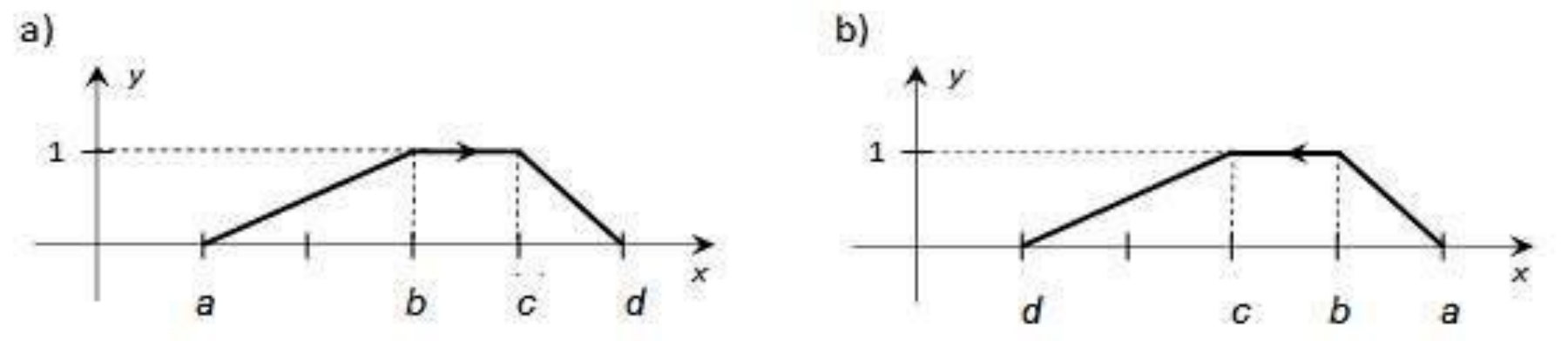

The membership function of TrOFN with:
a) positive orientation,
b) negative orientation.
Source: (Kosiński, 2006)

For the case $a = d$, TrOFN $\overleftrightarrow{Tr}(a,a,a,a)$ represents crisp number $a \in \mathbb{R}$, which is not oriented.

In agree with Kosiński's definitions of the arithmetic operations on OFN, the Kosiński's sum $\oplus$ of any TrOFN $\overleftrightarrow{Tr}(a_1,b_1,c_1,d_1)$ and $\overleftrightarrow{Tr}(a_2,b_2,c_2,d_2)$ is defined as follows:

$$\bar{\bar{C}}(a_1+a_2, b_1+b_2, c_1+c_2, d_1+d_2) = \overleftrightarrow{Tr}(a_1,b_1,c_1,d_1) \oplus \overleftrightarrow{Tr}(a_2,b_2,c_2,d_2) \qquad (4)$$

where the membership function $\mu_{Tr}(\cdot\,|a,b,c,d)$ of sum $\bar{\bar{C}}(a,b,c,d) \in \mathcal{F}(\mathbb{R})$ if is given by (3).

[3] Let us note that this identity describes additionally extended notation of numerical intervals, which is used in this work.

It is very easy to check that if TrOFN have identical orientations then results obtained by Kosiński's arithmetic are identical with results obtained by means of arithmetic introduced by Dubois and Prade (1978). Moreover, Kosiński (2006) has shown that if TrOFN have different orientations then results obtained by his arithmetic may be different to results obtained by arithmetic introduced by Dubois and Prade (1978).

**Counterexample:** Let us consider the Kosiński's sum $\bar{\bar{C}}$ of TrOFN $\overleftrightarrow{C} = \overleftrightarrow{Tr}(1;3;7;8)$ and $\overleftrightarrow{D} = \overleftrightarrow{Tr}(5;4;4;2)$ i.e.

$$\bar{\bar{C}}(6,7,11,10) = \bar{\bar{C}}(1+5,3+4,7+4,8+2) = \overleftrightarrow{Tr}(1;\ 3;\ 7;8) \oplus \overleftrightarrow{Tr}(5;4;4;2) \quad .$$

It is very easy to check, that the identity (3) determines such relation $\mu_{Tr}(\cdot\,|6,7,11,10)$ which is not any function. It implies that $\bar{\bar{C}}(6,7,11,10) \notin \mathcal{F}(\mathbb{R})$. Therefore, we say that The Kosiński's sum $\overleftrightarrow{Tr}(1;\ 3;\ 7;8) \oplus \overleftrightarrow{Tr}(5;4;4;2)$ does not exist. □

**Conclusion:** There exist such TrOFN that their Kosiński's sum does not exist.□

Therefore we ought to modify the operation of TrOFN adding in a way that the sum of TrOFN sum always exists. In this paper we define revised sum of TrOFN as follows

$$\overleftrightarrow{Tr}(a,b,c,d) \boxplus \overleftrightarrow{Tr}(p-a,q-b,r-c,s-d) =$$

$$= \begin{cases} \overleftrightarrow{Tr}(\min\{p,q\},q,r,\max\{r,s\}) & (q<r) \vee (q=r \wedge p \leq s) \\ \overleftrightarrow{Tr}(\max\{p,q\},q,r,\min\{r,s\}) & (q>r) \vee (q=r \wedge p>s) \end{cases} \tag{5}$$

For any pair of TrOFN their modified sum $\boxplus$ is equal to TrOFN determined such membership function that its graph is nearest to graph of relation determined by (3). It implies that if for fixed pair of TrOFN their Kosiński's sum $\oplus$ exists then it is equal to their revised sum $\boxplus$. If we find two different TrOFN nearest to considered Kosiński's sum $\oplus$ then we chose positively oriented one, which exists exactly one. It is very easy to check that revised sum (5) is commutative.

**Example 1:** Same cases of modified sum $\boxplus$ are presented bellow. Let us observe that in all these cases the Kosiński's sum $\oplus$ does not exist.

$$\overleftrightarrow{Tr}(1;\ 2;\ 4;6) \boxplus \overleftrightarrow{Tr}(5;3;2;1) = \overleftrightarrow{Tr}(5;5;6;7)$$

$$\overleftrightarrow{Tr}(6;\ 4;\ 2;1) \boxplus \overleftrightarrow{Tr}(1;2;3;5) = \overleftrightarrow{Tr}(7;6;5;5)$$

$$\overleftrightarrow{Tr}(1;\ 2;\ 4;4) \boxplus \overleftrightarrow{Tr}(5;3;2;1) = \overleftrightarrow{Tr}(5;5;6;6)$$

$$\overleftrightarrow{Tr}(4;\ 4;\ 2;1) \boxplus \overleftrightarrow{Tr}(1;2;3;5) = \overleftrightarrow{Tr}(6;6;5;5)$$

$$\overleftrightarrow{Tr}(1;\ 2;\ 3;4) \boxplus \overleftrightarrow{Tr}(6;3;2;2) = \overleftrightarrow{Tr}(7;5;5;5)$$

$$\overleftrightarrow{Tr}(1;\ 3;\ 7;8) \boxplus \overleftrightarrow{Tr}(5;4;4;2) = \overleftrightarrow{Tr}(6;7;11;11).$$

□

## 3. Associativity of the revised sum of ordered fuzzy numbers

For a more detailed view of further considerations, we distinguish the following special case of TrOFN.

**Definition 4:** For any monotonic sequence $\{a, b, c\} \subset \mathbb{R}$ triangular ordered fuzzy number (TOFN) $\overleftrightarrow{T}(a, b, c)$ is defined by the identity

$$\overleftrightarrow{T}(a, b, c) = \overleftrightarrow{Tr}(a, b, b, c)\ . \qquad (6)$$ □

Let us consider the following example.

**Example 2:** Let will be given to the following four TOFN:

$A = \overleftrightarrow{T}(10; 40; 70)$; $B = \overleftrightarrow{T}(110; 100; 60)$; $C = \overleftrightarrow{T}(50; 65; 105)$; $D = \overleftrightarrow{T}(120; 90; 67)$.

The number of different ways of associating three applications of the addition operator $\boxplus$ is equal the Catalan number $C_3 = 5$. Therefore, we have the following five different associations of four summands:

$(A \boxplus B) \boxplus (C \boxplus D)$; $\big((A \boxplus B) \boxplus C\big) \boxplus D$; $\big(A \boxplus (B \boxplus C)\big) \boxplus D$; $\big(A \boxplus (B \boxplus C)\big) \boxplus D$;

$A \boxplus \big(B \boxplus (C \boxplus D)\big)$ (Koshy, 2008).

For above expressions we obtain:

$$(A \boxplus B) \boxplus (C \boxplus D) = \left(\overleftrightarrow{T}(10; 40; 70) \boxplus \overleftrightarrow{T}(110; 100; 60)\right) \boxplus \left(\overleftrightarrow{T}(50; 65; 105) \boxplus \overleftrightarrow{T}(120; 90; 67)\right)$$

$$= \overleftrightarrow{T}(120; 140; 130) \boxplus \overleftrightarrow{T}(155; 155; 172) = \overleftrightarrow{T}(275; 295; 302)$$

$$\big((A \boxplus B) \boxplus C\big) \boxplus D = \left(\left(\overleftrightarrow{T}(10; 40; 70) \boxplus \overleftrightarrow{T}(110; 100; 60)\right) \boxplus \overleftrightarrow{T}(50; 65; 105)\right) \boxplus \overleftrightarrow{T}(120; 90; 67)$$
$$= \left(\overleftrightarrow{T}(120; 140; 140) \boxplus \overleftrightarrow{T}(50; 65; 105)\right) \boxplus \overleftrightarrow{T}(120; 90; 67)$$
$$= \overleftrightarrow{T}(170; 205; 245) \boxplus \overleftrightarrow{T}(120; 90; 67) = \overleftrightarrow{T}(290; 295; 312)$$

$$\big(A \boxplus (B \boxplus C)\big) \boxplus D = \left(\overleftrightarrow{T}(10; 40; 70) \boxplus \left(\overleftrightarrow{T}(110; 100; 60) \boxplus \overleftrightarrow{T}(50; 65; 105)\right)\right) \boxplus \overleftrightarrow{T}(120; 90; 67)$$
$$= \left(\overleftrightarrow{T}(10; 40; 70) \boxplus \overleftrightarrow{T}(160; 165; 165)\right) \boxplus \overleftrightarrow{T}(120; 90; 67)$$
$$= \overleftrightarrow{T}(170; 205; 235) \boxplus \overleftrightarrow{T}(120; 90; 67) = \overleftrightarrow{T}(290; 295; 302)$$

$$A \boxplus \big((B \boxplus C) \boxplus D\big) = \overleftrightarrow{T}(10; 40; 70) \boxplus \left(\left(\overleftrightarrow{T}(110; 100; 60) \boxplus \overleftrightarrow{T}(50; 65; 105)\right) \boxplus \overleftrightarrow{T}(120; 90; 67)\right)$$
$$= \overleftrightarrow{T}(10; 40; 70) \boxplus \left(\overleftrightarrow{T}(160; 165; 165) \boxplus \overleftrightarrow{T}(120; 90; 67)\right)$$
$$= \overleftrightarrow{T}(10; 40; 70) \boxplus \overleftrightarrow{T}(280; 255; 232) = \overleftrightarrow{T}(290; 295; 302)$$

$$A \boxplus \big(B \boxplus (C \boxplus D)\big) = \overleftrightarrow{T}(10; 40; 70) \boxplus \left(\overleftrightarrow{T}(110; 100; 60) \boxplus \left(\overleftrightarrow{T}(50; 65; 105) \boxplus \overleftrightarrow{T}(120; 90; 67)\right)\right)$$
$$= \overleftrightarrow{T}(10; 40; 70) \boxplus \left(\overleftrightarrow{T}(110; 100; 60) \boxplus \overleftrightarrow{T}(155; 155; 172)\right)$$
$$= \overleftrightarrow{T}(10; 40; 70) \boxplus \overleftrightarrow{T}(265; 255; 232) = \overleftrightarrow{T}(275; 295; 302)$$

In conclusion we see that we have here

$$(A \boxplus B) \boxplus (C \boxplus D) = A \boxplus \big(B \boxplus (C \boxplus D)\big) = \overleftrightarrow{T}(275; 295; 302)$$

$$\big(A \boxplus (B \boxplus C)\big) \boxplus D = A \boxplus \big((B \boxplus C) \boxplus D\big) = \overleftrightarrow{T}(290; 295; 302)$$

$$\big((A \boxplus B) \boxplus C\big) \boxplus D = \overleftrightarrow{T}(290; 295; 312). \square$$

The results of above example prove that revised sum $\boxplus$ of TrOFN is not associative.

## 4. The impact of summands permuting

Let us consider the following example.

**Example 3:** For all permutations of TOFN $A, B, C, D \in \mathbb{K}$ described in the Example 2, we determine their multiple sum:

$$\begin{aligned} A \boxplus B \boxplus C \boxplus D &= \overleftrightarrow{T}(10; 40; 70) \boxplus \overleftrightarrow{T}(110; 100; 60) \boxplus \overleftrightarrow{T}(50; 65; 105) \boxplus \overleftrightarrow{T}(120; 90; 67) \\ &= \overleftrightarrow{T}(120; 140; 140) \boxplus \overleftrightarrow{T}(50; 65; 105) \boxplus \overleftrightarrow{T}(120; 90; 67) \\ &= \overleftrightarrow{T}(170; 205; 245) \boxplus \overleftrightarrow{T}(120; 90; 67) = \overleftrightarrow{T}(290; 295; 312) \end{aligned}$$

$$\begin{aligned} A \boxplus B \boxplus D \boxplus C &= \overleftrightarrow{T}(10; 40; 70) \boxplus \overleftrightarrow{T}(110; 100; 60) \boxplus \overleftrightarrow{T}(120; 90; 67) \boxplus \overleftrightarrow{T}(50; 65; 105) \\ &= \overleftrightarrow{T}(120; 140; 140) \boxplus \overleftrightarrow{T}(120; 90; 67) \boxplus \overleftrightarrow{T}(50; 65; 105) \\ &= \overleftrightarrow{T}(240; 230; 207) \boxplus \overleftrightarrow{T}(50; 65; 105) = \overleftrightarrow{T}(290; 295; 312) \end{aligned}$$

$$\begin{aligned} A \boxplus C \boxplus B \boxplus D &= \overleftrightarrow{T}(10; 40; 70) \boxplus \overleftrightarrow{T}(50; 65; 105) \boxplus \overleftrightarrow{T}(110; 100; 60) \boxplus \overleftrightarrow{T}(120; 90; 67) \\ &= \overleftrightarrow{T}(60; 105; 175) \boxplus \overleftrightarrow{T}(110; 100; 60) \boxplus \overleftrightarrow{T}(120; 90; 67) \\ &= \overleftrightarrow{T}(170; 205; 235) \boxplus \overleftrightarrow{T}(120; 90; 67) = \overleftrightarrow{T}(290; 295; 302) \end{aligned}$$

$$\begin{aligned} A \boxplus C \boxplus D \boxplus B &= \overleftrightarrow{T}(10; 40; 70) \boxplus \overleftrightarrow{T}(50; 65; 105) \boxplus \overleftrightarrow{T}(120; 90; 67) \boxplus \overleftrightarrow{T}(110; 100; 60) \\ &= \overleftrightarrow{T}(60; 105; 175) \boxplus \overleftrightarrow{T}(120; 90; 67) \boxplus \overleftrightarrow{T}(110; 100; 60) \\ &= \overleftrightarrow{T}(180; 195; 242) \boxplus \overleftrightarrow{T}(110; 100; 60) = \overleftrightarrow{T}(290; 295; 302) \end{aligned}$$

$$\begin{aligned} A \boxplus D \boxplus B \boxplus C &= \overleftrightarrow{T}(10; 40; 70) \boxplus \overleftrightarrow{T}(120; 90; 67) \boxplus \overleftrightarrow{T}(110; 100; 60) \boxplus \overleftrightarrow{T}(50; 65; 105) \\ &= \overleftrightarrow{T}(130; 130; 137) \boxplus \overleftrightarrow{T}(110; 100; 60) \boxplus \overleftrightarrow{T}(50; 65; 105) \\ &= \overleftrightarrow{T}(240; 230; 197) \boxplus \overleftrightarrow{T}(50; 65; 105) = \overleftrightarrow{T}(290; 295; 302) \end{aligned}$$

$$\begin{aligned} A \boxplus D \boxplus C \boxplus B &= \overleftrightarrow{T}(10; 40; 70) \boxplus \overleftrightarrow{T}(120; 90; 67) \boxplus \overleftrightarrow{T}(50; 65; 105) \boxplus \overleftrightarrow{T}(110; 100; 60) \\ &= \overleftrightarrow{T}(130; 130; 137) \boxplus \overleftrightarrow{T}(50; 65; 105) \boxplus \overleftrightarrow{T}(110; 100; 60) \\ &= \overleftrightarrow{T}(180; 195; 237) \boxplus \overleftrightarrow{T}(110; 100; 60) = \overleftrightarrow{T}(290; 295; 302) \end{aligned}$$

$$\begin{aligned} B \boxplus A \boxplus C \boxplus D &= \overleftrightarrow{T}(110; 100; 60) \boxplus \overleftrightarrow{T}(10; 40; 70) \boxplus \overleftrightarrow{T}(50; 65; 105) \boxplus \overleftrightarrow{T}(120; 90; 67) \\ &= \overleftrightarrow{T}(120; 140; 140) \boxplus \overleftrightarrow{T}(50; 65; 105) \boxplus \overleftrightarrow{T}(120; 90; 67) \\ &= \overleftrightarrow{T}(170; 205; 245) \boxplus \overleftrightarrow{T}(120; 90; 67) = \overleftrightarrow{T}(290; 295; 312) \end{aligned}$$

$$\begin{aligned} B \boxplus A \boxplus D \boxplus C &= \overleftrightarrow{T}(110; 100; 60) \boxplus \overleftrightarrow{T}(10; 40; 70) \boxplus \overleftrightarrow{T}(120; 90; 67) \boxplus \overleftrightarrow{T}(50; 65; 105) \\ &= \overleftrightarrow{T}(120; 140; 140) \boxplus \overleftrightarrow{T}(120; 90; 67) \boxplus \overleftrightarrow{T}(50; 65; 105) \\ &= \overleftrightarrow{T}(240; 230; 207) \boxplus \overleftrightarrow{T}(50; 65; 105) = \overleftrightarrow{T}(290; 295; 312) \end{aligned}$$

$$
\begin{aligned}
B \boxplus C \boxplus A \boxplus D &= \overset{\leftrightarrows}{T}(110;100;60) \boxplus \overset{\leftrightarrows}{T}(50;65;105) \boxplus \overset{\leftrightarrows}{T}(10;40;70) \boxplus \overset{\leftrightarrows}{T}(120;90;67) \\
&= \overset{\leftrightarrows}{T}(160;165;165) \boxplus \overset{\leftrightarrows}{T}(10;40;70) \boxplus \overset{\leftrightarrows}{T}(120;90;67) \\
&= \overset{\leftrightarrows}{T}(170;205;235) \boxplus \overset{\leftrightarrows}{T}(120;90;67) = \overset{\leftrightarrows}{T}(290;295;302)
\end{aligned}
$$

$$
\begin{aligned}
B \boxplus C \boxplus D \boxplus A &= \overset{\leftrightarrows}{T}(110;100;60) \boxplus \overset{\leftrightarrows}{T}(50;65;105) \boxplus \overset{\leftrightarrows}{T}(120;90;67) \boxplus \overset{\leftrightarrows}{T}(10;40;70) \\
&= \overset{\leftrightarrows}{T}(160;165;165) \boxplus \overset{\leftrightarrows}{T}(120;90;67) \boxplus \overset{\leftrightarrows}{T}(10;40;70) \\
&= \overset{\leftrightarrows}{T}(280;255;232) \boxplus \overset{\leftrightarrows}{T}(10;40;70) = \overset{\leftrightarrows}{T}(290;295;302)
\end{aligned}
$$

$$
\begin{aligned}
B \boxplus D \boxplus A \boxplus C &= \overset{\leftrightarrows}{T}(110;100;60) \boxplus \overset{\leftrightarrows}{T}(120;90;67) \boxplus \overset{\leftrightarrows}{T}(10;40;70) \boxplus \overset{\leftrightarrows}{T}(50;65;105) \\
&= \overset{\leftrightarrows}{T}(230;190;127) \boxplus \overset{\leftrightarrows}{T}(10;40;70) \boxplus \overset{\leftrightarrows}{T}(50;65;105) \\
&= \overset{\leftrightarrows}{T}(230;190;127) \boxplus \overset{\leftrightarrows}{T}(10;40;70) \boxplus \overset{\leftrightarrows}{T}(50;65;105) \\
&= \overset{\leftrightarrows}{T}(240;230;197) \boxplus \overset{\leftrightarrows}{T}(50;65;105) = \overset{\leftrightarrows}{T}(290;295;302)
\end{aligned}
$$

$$
\begin{aligned}
B \boxplus D \boxplus C \boxplus A &= \overset{\leftrightarrows}{T}(110;100;60) \boxplus \overset{\leftrightarrows}{T}(120;90;67) \boxplus \overset{\leftrightarrows}{T}(50;65;105) \boxplus \overset{\leftrightarrows}{T}(10;40;70) \\
&= \overset{\leftrightarrows}{T}(230;190;127) \boxplus \overset{\leftrightarrows}{T}(50;65;105) \boxplus \overset{\leftrightarrows}{T}(10;40;70) \\
&= \overset{\leftrightarrows}{T}(230;190;127) \boxplus \overset{\leftrightarrows}{T}(50;65;105) \boxplus \overset{\leftrightarrows}{T}(10;40;70) \\
&= \overset{\leftrightarrows}{T}(280;255;232) \boxplus \overset{\leftrightarrows}{T}(10;40;70) = \overset{\leftrightarrows}{T}(290;295;302)
\end{aligned}
$$

$$
\begin{aligned}
C \boxplus A \boxplus B \boxplus D &= \overset{\leftrightarrows}{T}(50;65;105) \boxplus \overset{\leftrightarrows}{T}(10;40;70) \boxplus \overset{\leftrightarrows}{T}(110;100;60) \boxplus \overset{\leftrightarrows}{T}(120;90;67) \\
&= \overset{\leftrightarrows}{T}(60;105;175) \boxplus \overset{\leftrightarrows}{T}(110;100;60) \boxplus \overset{\leftrightarrows}{T}(120;90;67) \\
&= \overset{\leftrightarrows}{T}(170;205;235) \boxplus \overset{\leftrightarrows}{T}(120;90;67) = \overset{\leftrightarrows}{T}(290;295;302)
\end{aligned}
$$

$$
\begin{aligned}
C \boxplus A \boxplus D \boxplus B &= \overset{\leftrightarrows}{T}(50;65;105) \boxplus \overset{\leftrightarrows}{T}(10;40;70) \boxplus \overset{\leftrightarrows}{T}(120;90;67) \boxplus \overset{\leftrightarrows}{T}(110;100;60) \\
&= \overset{\leftrightarrows}{T}(60;105;175) \boxplus \overset{\leftrightarrows}{T}(120;90;67) \boxplus \overset{\leftrightarrows}{T}(110;100;60) \\
&= \overset{\leftrightarrows}{T}(180;195;242) \boxplus \overset{\leftrightarrows}{T}(110;100;60) = \overset{\leftrightarrows}{T}(290;295;302)
\end{aligned}
$$

$$
\begin{aligned}
C \boxplus B \boxplus A \boxplus D &= \overset{\leftrightarrows}{T}(50;65;105) \boxplus \overset{\leftrightarrows}{T}(110;100;60) \boxplus \overset{\leftrightarrows}{T}(10;40;70) \boxplus \overset{\leftrightarrows}{T}(120;90;67) \\
&= \overset{\leftrightarrows}{T}(160;165;165) \boxplus \overset{\leftrightarrows}{T}(10;40;70) \boxplus \overset{\leftrightarrows}{T}(120;90;67) \\
&= \overset{\leftrightarrows}{T}(170;205;235) \boxplus \overset{\leftrightarrows}{T}(120;90;67) = \overset{\leftrightarrows}{T}(290;295;302)
\end{aligned}
$$

$$
\begin{aligned}
C \boxplus B \boxplus D \boxplus A &= \overset{\leftrightarrows}{T}(50;65;105) \boxplus \overset{\leftrightarrows}{T}(110;100;60) \boxplus \overset{\leftrightarrows}{T}(120;90;67) \boxplus \overset{\leftrightarrows}{T}(10;40;70) \\
&= \overset{\leftrightarrows}{T}(160;165;165) \boxplus \overset{\leftrightarrows}{T}(120;90;67) \boxplus \overset{\leftrightarrows}{T}(10;40;70) \\
&= \overset{\leftrightarrows}{T}(280;255;232) \boxplus \overset{\leftrightarrows}{T}(10;40;70) = \overset{\leftrightarrows}{T}(290;295;302)
\end{aligned}
$$

$$
\begin{aligned}
C \boxplus D \boxplus A \boxplus B &= \overset{\leftrightarrows}{T}(50;65;105) \boxplus \overset{\leftrightarrows}{T}(120;90;67) \boxplus \overset{\leftrightarrows}{T}(10;40;70) \boxplus \overset{\leftrightarrows}{T}(110;100;60) \\
&= \overset{\leftrightarrows}{T}(155;155;172) \boxplus \overset{\leftrightarrows}{T}(10;40;70) \boxplus \overset{\leftrightarrows}{T}(110;100;60) \\
&= \overset{\leftrightarrows}{T}(165;195;242) \boxplus \overset{\leftrightarrows}{T}(110;100;60) = \overset{\leftrightarrows}{T}(275;295;302)
\end{aligned}
$$

$$
\begin{aligned}
C \boxplus D \boxplus B \boxplus A &= \overset{\leftrightarrows}{T}(50;65;105) \boxplus \overset{\leftrightarrows}{T}(120;90;67) \boxplus \overset{\leftrightarrows}{T}(110;100;60) \boxplus \overset{\leftrightarrows}{T}(10;40;70) \\
&= \overset{\leftrightarrows}{T}(155;155;172) \boxplus \overset{\leftrightarrows}{T}(110;100;60) \boxplus \overset{\leftrightarrows}{T}(10;40;70) \\
&= \overset{\leftrightarrows}{T}(265;255;232) \boxplus \overset{\leftrightarrows}{T}(10;40;70) = \overset{\leftrightarrows}{T}(275;295;302)
\end{aligned}
$$

$$
\begin{aligned}
D \boxplus A \boxplus B \boxplus C &= \overset{\leftrightarrows}{T}(120;90;67) \boxplus \overset{\leftrightarrows}{T}(10;40;70) \boxplus \overset{\leftrightarrows}{T}(110;100;60) \boxplus \overset{\leftrightarrows}{T}(50;65;105) \\
&= \overset{\leftrightarrows}{T}(130;130;137) \boxplus \overset{\leftrightarrows}{T}(110;100;60) \boxplus \overset{\leftrightarrows}{T}(50;65;105) \\
&= \overset{\leftrightarrows}{T}(240;230;197) \boxplus \overset{\leftrightarrows}{T}(50;65;105) = \overset{\leftrightarrows}{T}(290;295;302)
\end{aligned}
$$

$$
\begin{aligned}
D \boxplus A \boxplus C \boxplus B &= \overleftrightarrow{T}(120;90;67) \boxplus \overleftrightarrow{T}(10;40;70) \boxplus \overleftrightarrow{T}(50;65;105) \boxplus \overleftrightarrow{T}(110;100;60) \\
&= \overleftrightarrow{T}(130;130;137) \boxplus \overleftrightarrow{T}(50;65;105) \boxplus \overleftrightarrow{T}(110;100;60) \\
&= \overleftrightarrow{T}(180;195;142) \boxplus \overleftrightarrow{T}(110;100;60) = \overleftrightarrow{T}(290;295;302)
\end{aligned}
$$

$$
\begin{aligned}
D \boxplus B \boxplus A \boxplus C &= \overleftrightarrow{T}(120;90;67) \boxplus \overleftrightarrow{T}(110;100;60) \boxplus \overleftrightarrow{T}(10;40;70) \boxplus \overleftrightarrow{T}(50;65;105) \\
&= \overleftrightarrow{T}(230;190;127) \boxplus \overleftrightarrow{T}(10;40;70) \boxplus \overleftrightarrow{T}(50;65;105) \\
&= \overleftrightarrow{T}(240;230;197) \boxplus \overleftrightarrow{T}(50;65;105) = \overleftrightarrow{T}(290;295;302)
\end{aligned}
$$

$$
\begin{aligned}
D \boxplus B \boxplus C \boxplus A &= \overleftrightarrow{T}(120;90;67) \boxplus \overleftrightarrow{T}(110;100;60) \boxplus \overleftrightarrow{T}(50;65;105) \boxplus \overleftrightarrow{T}(10;40;70) \\
&= \overleftrightarrow{T}(230;190;127) \boxplus \overleftrightarrow{T}(50;65;105) \boxplus \overleftrightarrow{T}(10;40;70) \\
&= \overleftrightarrow{T}(280;255;232) \boxplus \overleftrightarrow{T}(10;40;70) = \overleftrightarrow{T}(290;295;302)
\end{aligned}
$$

$$
\begin{aligned}
D \boxplus C \boxplus A \boxplus B &= \overleftrightarrow{T}(120;90;67) \boxplus \overleftrightarrow{T}(50;65;105) \boxplus \overleftrightarrow{T}(10;40;70) \boxplus \overleftrightarrow{T}(110;100;60) \\
&= \overleftrightarrow{T}(155;155;172) \boxplus \overleftrightarrow{T}(10;40;70) \boxplus \overleftrightarrow{T}(110;100;60) \\
&= \overleftrightarrow{T}(165;195;242) \boxplus \overleftrightarrow{T}(110;100;60) = \overleftrightarrow{T}(275;295;302)
\end{aligned}
$$

$$
\begin{aligned}
D \boxplus C \boxplus B \boxplus A &= \overleftrightarrow{T}(120;90;67) \boxplus \overleftrightarrow{T}(50;65;105) \boxplus \overleftrightarrow{T}(110;100;60) \boxplus \overleftrightarrow{T}(10;40;70) \\
&= \overleftrightarrow{T}(155;155;172) \boxplus \overleftrightarrow{T}(110;100;60) \boxplus \overleftrightarrow{T}(10;40;70) \\
&= \overleftrightarrow{T}(265;255;232) \boxplus \overleftrightarrow{T}(10;40;70) = \overleftrightarrow{T}(275;295;302)
\end{aligned}
$$

In conclusion we see that we have here

$$
\begin{aligned}
C \boxplus D \boxplus A \boxplus B &= C \boxplus D \boxplus B \boxplus A = D \boxplus A \boxplus B \boxplus C = D \boxplus C \boxplus A \boxplus B = D \boxplus C \boxplus B \boxplus A \\
&= \overleftrightarrow{T}(275;295;302)
\end{aligned}
$$

$$
\begin{aligned}
A \boxplus C \boxplus B \boxplus D &= A \boxplus C \boxplus D \boxplus B = A \boxplus D \boxplus B \boxplus C = A \boxplus D \boxplus C \boxplus B = B \boxplus C \boxplus D \boxplus A \\
&= B \boxplus D \boxplus A \boxplus C = B \boxplus D \boxplus C \boxplus A = C \boxplus A \boxplus B \boxplus D = C \boxplus A \boxplus D \boxplus B \\
&= C \boxplus B \boxplus A \boxplus D = C \boxplus B \boxplus D \boxplus A = D \boxplus A \boxplus B \boxplus C = D \boxplus A \boxplus C \boxplus B \\
&= D \boxplus B \boxplus A \boxplus C = D \boxplus B \boxplus C \boxplus A = \overleftrightarrow{T}(290;295;302)
\end{aligned}
$$

$$
\begin{aligned}
A \boxplus B \boxplus C \boxplus D &= A \boxplus B \boxplus D \boxplus C = B \boxplus A \boxplus C \boxplus D = B \boxplus A \boxplus D \boxplus C = B \boxplus C \boxplus A \boxplus D \\
&= \overleftrightarrow{T}(290;295;312). \; \square
\end{aligned}
$$

The results of above example prove that multiple revised sum $\boxplus$ of finite sequence of TrOFN depends on its summands ordering.

## 5. Final conclusions

The summands ordering should be clearly defined for each practical application of the multiple revised sum $\boxplus$ of finite sequence of TrOFN. This summands ordering must be sufficiently justified in the field of application.